\documentclass[a4paper,12pt]{article}

\usepackage{mathtools}
\usepackage{graphicx}
\usepackage{physics}
\usepackage{subcaption}
\usepackage{amssymb}
\usepackage{relsize}
\usepackage{todonotes}
\usepackage{grffile}
\usepackage{cuted}
\usepackage{rsfso}
\usepackage[bottom]{footmisc}
\usepackage{placeins}
\usepackage{authblk}
\usepackage{cite}
\usepackage{hyperref}

\usepackage[a4paper,top=3cm,bottom=2cm,left=3cm,right=3cm,marginparwidth=1.75cm]{geometry}

\newcommand\myeq{\mathrel{\stackrel{\makebox[0pt]{\mbox{\normalfont\tiny\sffamily def}}}{=}}}
\newcommand{\Ncal}{{\cal N}}

\usepackage{xcolor}

\usepackage[%
]{pdfcomment}

\begin{document}

\title{Shaping the longitudinal intensity pattern of Cartesian beams in lossless and lossy media}

\author[1]{Mateus Corato-Zanarella\footnote{Corresponding author: mateuscorato@gmail.com}}
\author[2]{Henrique Corato-Zanarella}
\author[1]{Michel Zamboni-Rached}
\affil[1]{School of Electrical and Computer Engineering, University of Campinas, Campinas, SP, Brazil}
\affil[2]{Independent collaborator}

\date{}

\maketitle

\begin{abstract}

\par Several applications, such as optical tweezers and atom guiding, benefit from techniques that allow the engineering of spatial field profiles, in particular their longitudinal intensity patterns. In cylindrical coordinates, methods such as Frozen Waves allow an advanced control of beam characteristics, but in Cartesian coordinates there is no analogous technique. Since Cartesian beams may also be useful in applications, we develop here a method to modulate on demand the longitudinal intensity pattern of any (initially) unidimensional Cartesian beam with concentrated angular spectrum (thus encompassing all unidimensional paraxial beams) in lossless and lossy media. To this end, we write the total beam as a product of two unidimensional beams and explore the degree of freedom provided by the additional Cartesian coordinate. While in the plane where this coordinate is zero the chosen unidimensional beam keeps its structure with the additional desired intensity modulation, a sinusoidal-like oscillation appears in the direction of this variable and creates a spot whose size is tunable. Examples with Gaussian and Airy beams are presented and their corresponding experimental demonstrations in free-space are performed to show the validity of the method.

\end{abstract}

\section{Introduction}

\par The intensity profiles of optical beams are of fundamental importance for applications such as optical tweezers and atom guidance, as they directly impact the radiation forces. Consequently, engineering these patterns may open new possibilities for manipulating particles in efficient ways. In cylindrical coordinates, a powerful technique for controlling many fields properties is the Frozen Waves (FWs) method \cite{FW_1,FW_2,FW_3}, which uses superpositions of co-propagating Bessel beams. Since their phases depend linearly on the longitudinal coordinate $z$, their wavenumbers and complex amplitudes can be chosen based on truncated Fourier series to yield arbitrary longitudinal intensity patterns over the axis or cylindrical surfaces.

\par As unidimensional beams\footnote{Here, we say a beam is n-dimensional when it depends on n transverse spatial coordinates.}, written in Cartesian coordinates, may also have useful properties for particles manipulation, such as Gaussian beams \cite{GB_1,GB_2} and Airy beams \cite{airy_1D}, it could be useful to control their intensity patterns. The difficulty here is that their phases depend on $z$ in complicated ways, thus making it not possible to directly apply a method similar to that of FWs to cause the necessary interferences for arbitrarily modulating their longitudinal intensity patterns. However, if we make the resulting beam bidimensional by multiplying the profiles of two unidimensional beams, we can use the degree of freedom provided by the additional transverse coordinate to manipulate some features of the originally desired unidimensional beam.

\par In this work, we use the preceding idea to develop a method to arbitrarily modulate the longitudinal intensity pattern of unidimensional Cartesian beams with concentrated angular spectrum\footnote{It should be noted that all waves treated here are monochromatic and that we use the word spectrum to refer to the angular spectrum of plane waves, and not the frequency spectrum.}, thus encompassing, for example, all unidimensional paraxial beams. The technique can be applied to lossless and lossy media, allowing not only to compensate for the exponential decay due to absorption, but also to result in any desired longitudinal intensity pattern. To this end, inspired by the FWs method, we rely on a truncated Fourier series applied to the wavevector spectrum. If the unidimensional beam to be manipulated depends on $x$ and $z$, the resulting bidimensional beam reproduces its structure for $y=0$ with the additional longitudinal intensity modulation. Moreover, along the y-direction, a sinusoidal-like oscillation appears, introducing a spot around $y=0$ whose size can be chosen and, therefore, provides an additional transverse localization in this direction that may also be useful for applications. 

\par After the theoretical model is developed, we give an example in a lossy medium and present three sets of experimental data related to Gaussian and Airy beams that show the validity and feasibility of the proposed method. They exemplify interesting modulations that may be of practical interest, such as the compensation of punctual intensity diminishing due to diffraction, an exponentially-growing peak intensity and a rectangular longitudinal intensity pattern, the last of which illustrates how a beam can be ``turned on'' or ``turned off'' over regions on demand, making it possible for it to selectively interact with particles at specific positions while avoiding others, for instance.

\par As our proposed method emphasizes localization effects, its methodology could be extended to pulses and have interesting implications in the propagation of light in dispersive disordered media \cite{dispersive_disordered_media}. In addition, as our technique allows high energy concentration within spatial regions chosen on demand, it may be used to explore nonlinear phenomena, such as optical spatial solitons, in material media \cite{nonlinear}.

\section{Theoretical model}
\label{theory}

\par Any monochromatic wave $\Psi(x,y,z,t)$, which is a solution of the homogeneous Helmholtz equation $(\nabla^2+k^2)\Psi=0$, can be written as a superposition of plane waves with the same angular frequency $\omega$ and different wavevectors $\vec{k}=k_x\hat{x}+k_y\hat{y}+k_z\hat{z}=k\hat{k}$, given that $k^2=k_x^2+k_y^2+k_z^2$ and $k=k_r+ik_i=n\omega/c$, where $n=n_r+in_i$ is the complex refractive index of the medium. We will consider that both $k_x$ and $k_y$ are real, as if the wave was generated in a lossless material before penetrating into the absorbing medium, with the interface between the media being perpendicular to the z-direction. If we consider only waves propagating in the $+\hat{z}$ direction, $k_z=\sqrt{k^2-(k_x^2+k_y^2)}$ and $\Psi(x,y,z,t)$ can be written as\footnote{The integration is performed within a region that excludes evanescent waves when $n_i \rightarrow 0$.}
\begin{equation}
\Psi(x,y,z,t)=e^{-i\omega t} \iint\limits_{k_x^2+k_y^2\leq k_r^2} \text{d}k_x\text{d}k_y \, \tilde{S}(k_x,k_y) e^{ik_x x} e^{ik_y y} e^{i \sqrt{k^2-(k_x^2+k_y^2)}z}
\label{superposition_of_plane_waves}
\end{equation}

\noindent where $\tilde{S}(k_x,k_y)$ is the spectrum, which determines the amplitude and phase of each plane wave. Even if $\tilde{S}(k_x,k_y)$ is separable in $k_x$ and $k_y$, i.e., $\tilde{S}(k_x,k_y)=\tilde{S}_x(k_x)\tilde{S}_y(k_y)$, it is generally not possible to find solutions of Eq. \eqref{superposition_of_plane_waves} of the form $\Psi(x,y,z,t)=\mathbb{X}(x,z,t)\mathbb{Y}(y,z,t)$, since $k_z=\sqrt{k^2-(k_x^2+k_y^2)}$ couples the integrals. However, if the spectrum is concentrated around $(k_x,k_y)=(k_{x_0},k_{y_0})$, some simplifications can be made. Mathematically, it means that $|\tilde{S}(k_x,k_y,\omega)|$ is significant only where $|k_x^{\prime}/k_{x_0}| \ll 1$ and $|k_y^{\prime}/k_{y_0}| \ll 1$, with $k_x^{\prime}\myeq k_x-k_{x_0}$ and $k_y^{\prime}\myeq k_y-k_{y_0}$ being defined as the deviation variables. In this region, a Taylor series expansion can be used to approximate $k_z$ to second order in $k_x^\prime$ and $k_y^\prime$, resulting in
\begin{equation}
k_z \approx k_{z_0} -a_x k^{\prime}_x -a_y k^{\prime}_y -b{k^{\prime}_x}^2 - b{k^{\prime}_y}^2 - d{k^{\prime}_x}{k^{\prime}_y}
\label{kz_approx}
\end{equation}

\noindent where $k_{z_0} \myeq \sqrt{k^2-k_{\mathsmaller{\perp}_0}^2}=k_{z_{0_R}}+ik_{z_{0_I}}$, $k_{\mathsmaller{\perp}_0}^2\myeq k_{y_0}^2+k_{x_0}^2$ and

\begin{subequations}
	\begin{equation}
	a_x\myeq\frac{k_{x_0}}{\sqrt{k^2-k_{\mathsmaller{\perp}_0}^2}} \,\text{,} \quad a_y\myeq\frac{k_{y_0}}{\sqrt{k^2-k_{\mathsmaller{\perp_0}}^2}}
	\end{equation}
	\begin{equation}
	b \myeq \frac{1}{2}\frac{k^2}{\left(k^2-k_{\mathsmaller{\perp_0}}^2\right)^{\frac{3}{2}}}
	\end{equation}
	\begin{equation}
	d \myeq \frac{k_{x_0}k_{y_0}}{\left(k^2-k_{\mathsmaller{\perp_0}}^2\right)^{\frac{3}{2}}}
	\end{equation}
	\label{constants_in_initial_approximation}
\end{subequations}

\noindent are complex constants.
\par Changing the integration variables in Eq. \eqref{superposition_of_plane_waves} to $k^{\prime}_x$ and $k^{\prime}_y$ and noting that the limits can be extended to $-\infty$ and $\infty$ without loss of accuracy due to the fact that the spectrum is significant only around $(k^{\prime}_x,k^{\prime}_y)=(0,0)$, we get
\begin{equation}
\Psi(x,y,z,t) \approx e^{-i\omega t} e^{ik_{x_0}x} e^{ik_{y_0}y} e^{ik_{z_0}z} A(x,y,z)
\end{equation}

\noindent with
\begin{align}
A(x,y,z) = \int_{-\infty}^{+\infty} \mathrm{d}k_x^{\mathsmaller\prime} \int_{-\infty}^{+\infty} \mathrm{d}k_y^{\mathsmaller\prime} S(k_x^{\mathsmaller\prime} , k_y^{\mathsmaller\prime}) e^{ik_x^{\mathsmaller\prime}x} e^{ik_y^{\mathsmaller\prime}y} e^{-i\left(a_xk_x^{\mathsmaller\prime}+a_yk_y^{\mathsmaller\prime}+bk_x^{\mathsmaller\prime \, 2}+bk_y^{\mathsmaller\prime \, 2} +dk_x^{\mathsmaller\prime} k_y^{\mathsmaller\prime}\right)z} 
\end{align}

\noindent where $S(k^{\prime}_x,k^{\prime}_y)\myeq \tilde{S}(k_{x_0}+k^{\prime}_x,k_{y_0}+k^{\prime}_y)$.

\par Even with this approximation, $e^{-idk_x^{\mathsmaller\prime} k_y^{\mathsmaller\prime}z}$ still couples the integrals. However, in the case either $k_{x_0}$ or $k_{y_0}$ is zero, $d=0$ and the coupling vanishes. In these cases, or even when this coupling term may be neglected compared to the others, if we choose a separable spectrum $S(k^{\prime}_x,k^{\prime}_y)=S_x(k^{\prime}_x)S_y(k^{\prime}_y)$, $A(x,y,z)$ can be written as a product of two functions, one dependent on $x$ and $z$ and another on $y$ and $z$:
\begin{subequations}
	\begin{align}
	&A(x,y,z) = \chi(x,z)\gamma(y,z)
	\label{separation_of_A} \\
	&\chi(x,z) = \int_{-\infty}^{+\infty}\mathrm{d}k_x^{\mathsmaller\prime}S_x(k_x^{\mathsmaller\prime})e^{ik_x^{\mathsmaller\prime}x}e^{-ia_xk_x^{\mathsmaller\prime}z}e^{-ibk_x^{\mathsmaller\prime \, 2}z}
	\label{expression_of_phi} \\
	&\gamma(y,z) = \int_{-\infty}^{+\infty}\mathrm{d}k_y^{\mathsmaller\prime}S_y(k_y^{\mathsmaller\prime})e^{ik_y^{\mathsmaller\prime}y}e^{-ia_yk_y^{\mathsmaller\prime}z}e^{-ibk_y^{\mathsmaller\prime \, 2}z}
	\label{expression_of_xi}
	\end{align}
\end{subequations}

\noindent that is, $A(x,y,z)$ is a product of the envelopes $\chi(x,z)$ and $\gamma(y,z)$ of two unidimensional beams, as long as their spectra $S_x(k^\prime_x)$ and $S_y(k^\prime_y)$ are concentrated around $k^\prime_x=0$ and $k^\prime_y=0$.

\par Our goal here is to somehow use $\gamma(y,z)$ to control the intensity of a desired unidimensional beam with envelope $\chi(x,z)$ along its propagation inside the interval $0\leq z \leq L$, modulating its usual intensity by an arbitrary function $|F(z)|^2$. To this end, we should have $|\gamma(0,z)|=\left|e^{k_{z_{0_I}}z}F(z)\right|$, so that $|\Psi(x,0,z,t)|^2=\left|e^{-k_{z_{0_I}}z}A(x,0,z)\right|^2=|F(z)|^2|\chi(x,z)|^2$. In order to accomplish this, we use the freedom for choosing $S_y(k^\prime_y)$. At $y=0$, Eq. \eqref{expression_of_xi} becomes
\begin{equation}
\gamma(0,z) = \int_{-\infty}^{+\infty}\mathrm{d}k_y^{\mathsmaller\prime}S_y(k_y^{\mathsmaller\prime})e^{-i(a_yk_y^{\mathsmaller\prime}+bk_y^{\mathsmaller\prime \, 2})z}
\label{xi,y=0}
\end{equation}

\noindent which resembles an inverse Fourier transform. To make the analysis analytically simpler, we will turn the integral into a discrete sum by choosing
\begin{equation}
S_y(k_y^{\mathsmaller\prime}) = \sum_{m = -M}^{M}A_M 	\delta\left(k_y^{\mathsmaller\prime} - k_{y_m}^{\mathsmaller\prime}\right)
\label{discrete_spectrum}
\end{equation}

\noindent where $\delta(\cdot)$ is the Dirac delta function. Then, Eq. \eqref{xi,y=0} becomes
\begin{align}
\nonumber \gamma(0,z) = &\sum_{m = -M}^{M}A_m e^{-i\left(a_y k_{y_m}^{\mathsmaller\prime} + bk_{y_m}^{\mathsmaller\prime \, 2}   \right)z}\\
= &\sum_{m = -M}^{M}A_m e^{-i\left(a_{y_R} k_{y_m}^{\mathsmaller\prime} + b_R k_{y_m}^{\mathsmaller\prime \, 2}\right)z}
e^{-\left(a_{y_I} k_{y_m}^{\mathsmaller\prime} + b_I k_{y_m}^{\mathsmaller\prime \, 2}\right)z}
\label{xi(0,z)_with_discrete_spectrum_original}
\end{align}

\noindent where $a_y$ and $b$ where expanded into their real and imaginary parts, i.e., $a_y=a_{y_R}-ia_{y_I}$ and $b=b_R-ib_I$. Since the most significant term in the superposition usually corresponds to $m=0$, as pointed out later in this section, and $k_{y_m}^{\mathsmaller\prime}$ is close to $k_{y_0}^{\mathsmaller\prime}$, the exponentially-decaying factor for any $m$ can be approximated by its value for $m=0$, resulting in
\begin{equation}
\gamma(0,z) \approx e^{-\left(a_{y_I} k_{y_0}^{\mathsmaller\prime} + b_I k_{y_0}^{\mathsmaller\prime \, 2}\right)z}
\sum_{m = -M}^{M}A_m e^{-i\left(a_{y_R} k_{y_m}^{\mathsmaller\prime} + b_R k_{y_m}^{\mathsmaller\prime \, 2}\right)z}
\label{xi(0,z)_with_discrete_spectrum}
\end{equation}

\noindent which now resembles a truncated Fourier series. This is indeed the case if we make the choice
\begin{equation}
-(a_{y_R}k_{y_m}^{\mathsmaller\prime} + b_R k_{y_m}^{\mathsmaller\prime \, 2}) = \frac{2\pi m}{L} - Q\,\text{,}\qquad Q\in\mathbb{R}^{+}
\label{defining_expression_for_k_{y_n}}	
\end{equation}

\noindent which transforms Eq. \eqref{xi(0,z)_with_discrete_spectrum} into
\begin{equation}
\gamma(0,z) \approx e^{-iQz} e^{-\left(a_{y_I} k_{y_0}^{\mathsmaller\prime} + b_I k_{y_0}^{\mathsmaller\prime \, 2}\right)z} \sum_{m = -M}^{M}A_m\exp\left[i\frac{2\pi m}{L}z\right]
\label{xi_with_summation}
\end{equation}

\par Then, if we choose the coefficients $A_m$ according to
\begin{equation}
A_m = \frac{1}{L}\int_{0}^{L} F(z)e^{k_{z_{0_I}}z} \exp\left[\left(a_{y_I} k_{y_0}^{\mathsmaller\prime} + b_I k_{y_0}^{\mathsmaller\prime \, 2}\right)z\right] e^{-i\frac{2\pi m}{L}z}\,\mathrm{d}z
\label{definition_of_A_n}
\end{equation}

\noindent we obtain $\gamma(0,z)\approx e^{-iQz}e^{k_{z_{0_I}}z}F(z)$ and the desired relation $|\gamma(0,z)|=\left|e^{k_{z_{0_I}}z}F(z)\right|$. Of course, $\chi(x,z)$ will also have a decay due to absorption in the case of lossy medium, but it can be compensated when defining $F(z)$. Notice that if $\left|k_{\perp_0}/k\right|$ is small, the absorption will lie almost entirely in the factor $e^{-k_{z_{0_I}}z}$.

\par Solving Eq. \eqref{defining_expression_for_k_{y_n}}	gives the $k_{y_m}^{\mathsmaller\prime}$ that should be used in the superposition:
\begin{equation}
k_{y_m}^{\mathsmaller\prime} = -\frac{a_{y_R}}{2b_R} \, \pm \,\sqrt{\left(\frac{a_{y_R}}{2b_R}\right)^2 + \frac{1}{b_R}\left(Q - \frac{2\pi m}{L}\right)}
\label{k_y_n_prime_isolated}
\end{equation}

\noindent where any of the signs can be chosen. 

\par So far, we have focused at $y=0$, where the intensity modulation has been done. To comprehend what happens for $y\neq 0$, we have to qualitatively understand the meaning of $e^{ik_{y_0}}\gamma(y,z)$. According to Eq. \eqref{expression_of_xi}, it is a superposition of plane waves with wavenumbers in the y-direction that slightly vary around $k_{y_0}$, so that they all carry power to the $+\hat{y}$ direction if $k_{y_0}>0$ or to the $-\hat{y}$ direction if $k_{y_0}<0$. All the energy necessary for the modulation at $y=0$ comes from this flux, which requires a sufficiently intense initial lateral structure within the region $y<0$ if $k_{y_0}>0$ or $y>0$ if $k_{y_0}<0$. As a consequence, the beam does not have a good transverse localization along the y-direction. In the case of lossy media, the lateral structure has to be more intense than in the lossless case, since part of the energy is absorbed when it moves towards the center of the beam. The higher the losses and the modulation interval are, the more intense the lateral structure will have to be.

\par To overcome the localization issue, we may improve our model to symmetrically combine beams propagating in opposite directions, that is
\begin{align}
\Psi(x,y,z,t) \approx  e^{-i\omega t}e^{ik_{x_0}x}e^{ik_{z_0}z} \frac{\left[e^{ik_{y_0}y}\gamma_1(y,z) + e^{-ik_{y_0}y}\gamma_2(y,z)\right]}{2}\chi(x,z)
\label{Psi_with_symmetric_sum}
\end{align}

\noindent where
\begin{subequations}
	\begin{align}
	\nonumber \gamma_p(y,z) = e^{-iQz} & \sum_{m = -M}^{M}A_m \exp\left[i\frac{2\pi m}{L}z\right] \\
	\nonumber & \times \exp\left[-\left((-1)^{p-1}a_{y_I} k_{y_{m,p}}^{\mathsmaller\prime} + b_I k_{y_{m,p}}^{\mathsmaller\prime\,2}\right)z\right]\\
	& \times \exp\left[ik_{y_{m,p}}^{\mathsmaller\prime}y\right]\,\text{,}\quad p=1,2
	\label{xi_with_symmetric_summation}
	\end{align}
	\begin{equation}
	k_{y_{m,1}}^{\mathsmaller\prime} = -\frac{a_{y_R}}{2b_R} \, \pm \,\sqrt{\left(\frac{a_{y_R}}{2b_R}\right)^2 + \frac{1}{b_R}\left(Q - \frac{2\pi m}{L}\right)}
	\label{k_y_n_1_prime}
	\end{equation}
	\begin{equation}
	k_{y_{m,2}}^{\mathsmaller\prime} =-k_{y_{m,1}}^{\mathsmaller\prime}= \frac{a_{y_R}}{2b_R} \, \mp \,\sqrt{\left(\frac{a_{y_R}}{2b_R}\right)^2 + \frac{1}{b_R}\left(Q - \frac{2\pi m}{L}\right)}
	\label{k_y_n_2_prime}
	\end{equation}
	\label{k_y_n_and_more}
\end{subequations}	

\noindent and $a_y$ was defined with $k_{y_0}$\footnote{This is the reason for the $(-1)^{p-1}$ factor in Eq. \eqref{xi_with_symmetric_summation}.}. In this way, there are waves coming from both directions that contribute to the modulation at $y=0$, thus making the lateral structure of the beam symmetric and decreasing its overall intensity. Also, their interference creates a pattern with oscillations in the y-direction that define a spot around $y=0$, which may be useful in practice. To visualize this, we can rearrange the y-dependent terms in Eq. \eqref{Psi_with_symmetric_sum} to result in
\begin{align}
\nonumber & \Psi(x,y,z,t) \approx e^{-i\omega t}e^{ik_{x_0}x}e^{ik_{z_0}z} e^{-iQz} \chi(x,z)  \\
\nonumber & \times \sum_{m = -M}^{M}A_m \exp\left[i\frac{2\pi m}{L}z\right] \exp\left[-\left(a_{y_I} k_{y_{m,1}}^{\mathsmaller\prime} + b_I k_{y_{m,1}}^{\mathsmaller\prime\,2}\right)z\right]\\
&\times \cos\left[\left(k_{y_0} - \frac{a_{y_R}}{2b_R} \pm \sqrt{\left(\frac{a_{y_R}}{2b_R}\right)^2 + \frac{1}{b_R}\left(Q - \frac{2\pi m}{L}\right)}\right)y\right]
\label{psi_cos}
\end{align}

\par Note that the resulting $\tilde{S}_y(k_y)$ of this beam has peaks concentrated around both $k_{y_0}$ and $-k_{y_0}$.

\par Usually, the most significant coefficient in Eq. \eqref{psi_cos} is $A_0$ and $|A_m|$ gets progressively less important when $|m|$ increases. This is due to the properties of the Fourier coefficients and is consistent with our hypothesis that the spectra of $\gamma_p(y,z)$ ($p=1,2$) are concentrated. Consequently, the main contribution in Eq. \eqref{psi_cos} comes from the cosine term with $m=0$ and we can use it to estimate the resulting spot radius. It has the form $\cos(k_{y_c}y)$ with
\begin{equation}
k_{y_c}\myeq k_{y_0} - \frac{a_{y_R}}{2b_R} \pm \sqrt{\left(\frac{a_{y_R}}{2b_R}\right)^2 + \frac{Q}{b_R}}
\label{k_y_c}
\end{equation}

\noindent and has a period of $Y=2\pi/|k_{y_c}|$. The spot radius is then 
\begin{equation}
\Delta y=\frac{Y}{4}=\frac{\pi}{2|k_{y_c}|}=\frac{\pi}{2\abs{k_{y_0} - \frac{a_{y_R}}{2b_R} \pm \sqrt{\left(\frac{a_{y_R}}{2b_R}\right)^2 + \frac{Q}{b_R}}}}
\label{spot}
\end{equation}

\par Note that $\Delta y$ is defined by $k_{y_0}$ and $Q$. However, $Q$ also plays another very important role, as it is necessary to ensure that every $k_{y_m}=k_{y_0}+k_{y_m}^{\mathsmaller\prime}$ is real. If $k_{y_m}$ had a non-zero imaginary part, its corresponding plane wave in the superposition would have an evanescent behavior in the y-direction, with a factor of the form $e^{\pm \alpha y}$, where $\alpha \in \mathbb{R}^+$. Either of these signs would imply $|\Psi(x,y,z,t)| \rightarrow \infty$ as either $y \rightarrow \infty$ or $y\rightarrow -\infty$. Therefore, imposing the term inside the square root in Eq. \eqref{k_y_n_prime_isolated} to be positive for every $m$, we derive that $Q$ must satisfy the constraint
\begin{equation}
Q \geq \frac{2\pi M}{L} - \frac{a_{y_R}^2}{4b_R}
\label{condition1}
\end{equation}

\par The initial choice of $Q\geq 0$ in Eq. \eqref{defining_expression_for_k_{y_n}} is due to the same reason. In this expression, if $a_{y_R}=0$ and $m=0$, we would have $k_{y_0}^{\mathsmaller\prime\,2}=Q/b_R$ and, since $b_R>0$, a negative $Q$ would imply an imaginary $k_{y_0}^{\mathsmaller\prime}$.

\par In addition, for all the waves in the superposition to be propagating in the z-direction when $n_i \rightarrow 0$, we must have
\begin{equation}
\max_m\left[k_{y_m}^2\right]+\max\left[k_x^2\right]\leq k_r^2
\label{condition2}
\end{equation}

\noindent which is also a constraint on the values of $k_{y_0}$, $Q$, $M$ and $L$.

\par It is worth mentioning that, according to Eq. \eqref{k_y_c}, $k_{y_c}$ and $-k_{y_c}$, which are where $\tilde{S}_y(k_y)$ usually has peaks, are not equal to $k_{y_0}$ and $-k_{y_0}$, around which the series expansion of Eq. \eqref{kz_approx} was made. However, this is not an issue, as $k_{y_c}$ and $k_{y_0}$ are close and, therefore, $|\tilde{S}_y(k_y)|$ is significant only around $k_{y_0}$ and $-k_{y_0}$, as assumed.

\par Although it may seem that the aforementioned constraints are limiting, in fact they are not and we still have plenty of freedom to choose the parameters for $\gamma(y,z)$. A good choice for $k_{y_0}$ is $k_{y_0}=0$, which means that $\gamma(y,z)$ is a paraxial beam. In this way, $d=0$ for any $k_{x_0}$, so that the hypothesis of a separable $A(x,y,z)$ is always satisfied for any $\chi(x,z)$ with concentrated $\tilde{S}_x(k_x)$. In addition, the expressions become a lot simpler and, as we will show soon, the possible ranges of values for $\Delta y$, $L$ and $M$ are reasonable for practical applications.

\par For $k_{y_0}=0$, Eq. \eqref{psi_cos} to Eq. \eqref{condition2} become
\begin{align}
\nonumber &\Psi(x,y,z,t) = e^{-i\omega t}e^{ik_{x_0}x}e^{ik_{z_0}z}e^{-iQz}\chi(x,z) \\
\nonumber & \times \sum_{m = -M}^{M}A_m \exp\left[i\frac{2\pi m}{L}z\right] \exp\left[-\left(a_{yI} k_{y_{m,1}}^{\mathsmaller\prime} + b_I k_{y_{m,1}}^{\mathsmaller\prime\,2}\right)z\right] \\
& \times \cos(\sqrt{\frac{1}{b_R}\left(Q-\frac{2\pi m}{L}\right)y})
\label{psi_cos_paraxial}
\end{align}
\begin{equation}
k_{y_c}=\sqrt{\frac{Q}{b_R}}
\label{kyc_paraxial}
\end{equation}
\begin{equation}
\Delta y = \frac{\pi}{2}\sqrt{\frac{b_R}{Q}}
\label{spot_paraxial}
\end{equation}
\begin{equation}
Q \geq \frac{2\pi M}{L}
\label{Q_paraxial}
\end{equation}
\begin{equation}
\frac{1}{b_R}\left(Q +\frac{2 \pi M}{L} \right) +\max\left[k_x^2\right] \leq k_r^2
\end{equation}

\par In this case, the two peaks of $\tilde{S}_y(k_y)$ are symmetrically displaced around $k_{y_0}=0$ at the positions $k_y=\pm k_{y_c}$. Thus, as long as $|k_{y_c}|$ is not high enough to make $\gamma(y,z)$ nonparaxial, the model with $k_{y_0}=0$ is still valid and, based on it, we can make some estimates. If we take $k_{y_c}/k_r=0.1$ as a reasonable heuristic criterion for an upper limit, Eq. \eqref{spot_paraxial} and Eq. \eqref{kyc_paraxial} tell us that the spot may be as low as $\Delta y_{\text{min}} \sim 2.5 \lambda$, while in any case it can be as high as $\Delta y_{\text{max}}=\frac{\pi}{2}\sqrt{\frac{b_RL}{2\pi M}}$ (assuming $Q$ at its minimum possible value of $Q=2\pi M/L$). Considering that a $M$ of a few tens is usually enough to provide a good approximation for $F(z)$, the application of the criterion $k_{y_c}/k_r\leq 0.1$ with $Q=2\pi M/L$ to the common case of paraxial $\chi(x,z)$ in low-loss medium (for which $b_R\approx 1/(2k_R)$) implies that $L$ should be, in principle, of at least a few thousands of $\lambda$. However, as we will show in the examples, the interval in which the field is to be modeled does not need to be the whole interval $0\leq z \leq L$, i.e., it can be $0\leq z \leq L_d$ with $L_d<L$. Although in this situation a higher $M$ is necessary for the truncated Fourier series to approximate $F(z)$ with a given precision, it is possible to tailor $L$ and $M$ to suit our needs.

\par It should be noted that, since $\Psi(x,y,z,t)$ was obtained using a truncated Fourier series in the z-direction, its longitudinal intensity pattern is periodic with period $L$. As a result, the beam transverse pattern at $z=0$ has repetitions of more or less the same structure along the y-direction and each of them is responsible for the modulation of the field in one of the periods. Experimentally, this periodicity can be eliminated by filtering the structures responsible for the periods beyond the first. To this end, it is important to estimate the length of the lateral structure responsible for the first period. To accomplish the modulation at each desired longitudinal position, there must be energy arriving from the sides. Noting that the wave's velocity in the $y$ and $z$ directions are given respectively by $vk_{y_c}/k_r$ and $vk_{z_{0_R}}/k_r$ (where $v=n_r\omega/c$) and equating the time it takes for the wave to propagate a distance $L_d$ to the time it takes for the outer portion of the lateral structure (located at $y=L_y$) to arrive at $y=0$, we get
\begin{equation}
L_y=L_d\frac{k_{y_c}}{k_{z_{0_R}}}
\label{Ly}
\end{equation}

\noindent where $L_y$ denotes half of the total length of the lateral structure responsible for the first period.

\section{Example in a lossy medium}
\label{sec:perdas}

\par To illustrate the proposed method, we present here a theoretical example of an initially unidimensional paraxial Gaussian beam with spot size $w_0=50\,\mu\text{m}$ modulated to have a constant amplitude at $x=y=0$ inside an absorbing material with $n=2+i7.5\times 10^{-6}$ at $\lambda=632.8\,\text{nm}$, implying a penetration depth of $Z=1/(2k_i)\approx 0.67\,\text{cm}$ for usual beams. To show how our technique can be used to compensate the exponential decay, we choose $L_d=4Z\approx 2.68\,\text{cm}$. Also, we adopt $M=25$ and $L=4L_d$, resulting in a spot size ($\Delta y\approx 6.52\,\mu\text{m}$) and in a lateral structure ($L_y\approx 326\,\mu\text{m}$) that could both be visible in the same figure. Although already pointed out in sec. \ref{theory}, the difference between $L$ and $L_d$ will be explained in more detail in sec. \ref{experiment}.

\par As a reference, the expression for the envelope of an unidimensional Gaussian beam is \cite{GB_unidimensional}
\begin{equation}
A_{GB}(x,z)= \frac{C}{\sqrt{1 + i\frac{2}{k w_0^2}z}} \exp \left[-\frac{x^2}{w_0^2 \left(1 + i\frac{2}{k w_0^2}z \right)} \right]
\label{gaussian_envelope}
\end{equation}

\noindent where $C$ is a constant. In this case, $F(z)$ was chosen so to compensate the natural diminishing of the intensity of the unidimensional Gaussian beam at $x = y = 0$ due to diffraction. Since, according to Eq. \eqref{gaussian_envelope}, its amplitude at $x=0$ is
\begin{equation}
\abs{A_{GB}(0,z)} = \frac{|C|}{\left(1 + \frac{4}{k^2 w_0^4}z^2 \right)^{1/4}}
\end{equation}

\noindent $F(z)$ was chosen as
\begin{equation}
F(z) = \left(1 + \frac{4}{k^2 w_0^4}z^2 \right)^{1/4}
\label{F_for_gaussian}
\end{equation}

\noindent so that $|F(z)A(0,z)|=|C|$.

\par
\begin{figure} [htbp]
	\centering
	\includegraphics[width = 0.7\columnwidth]{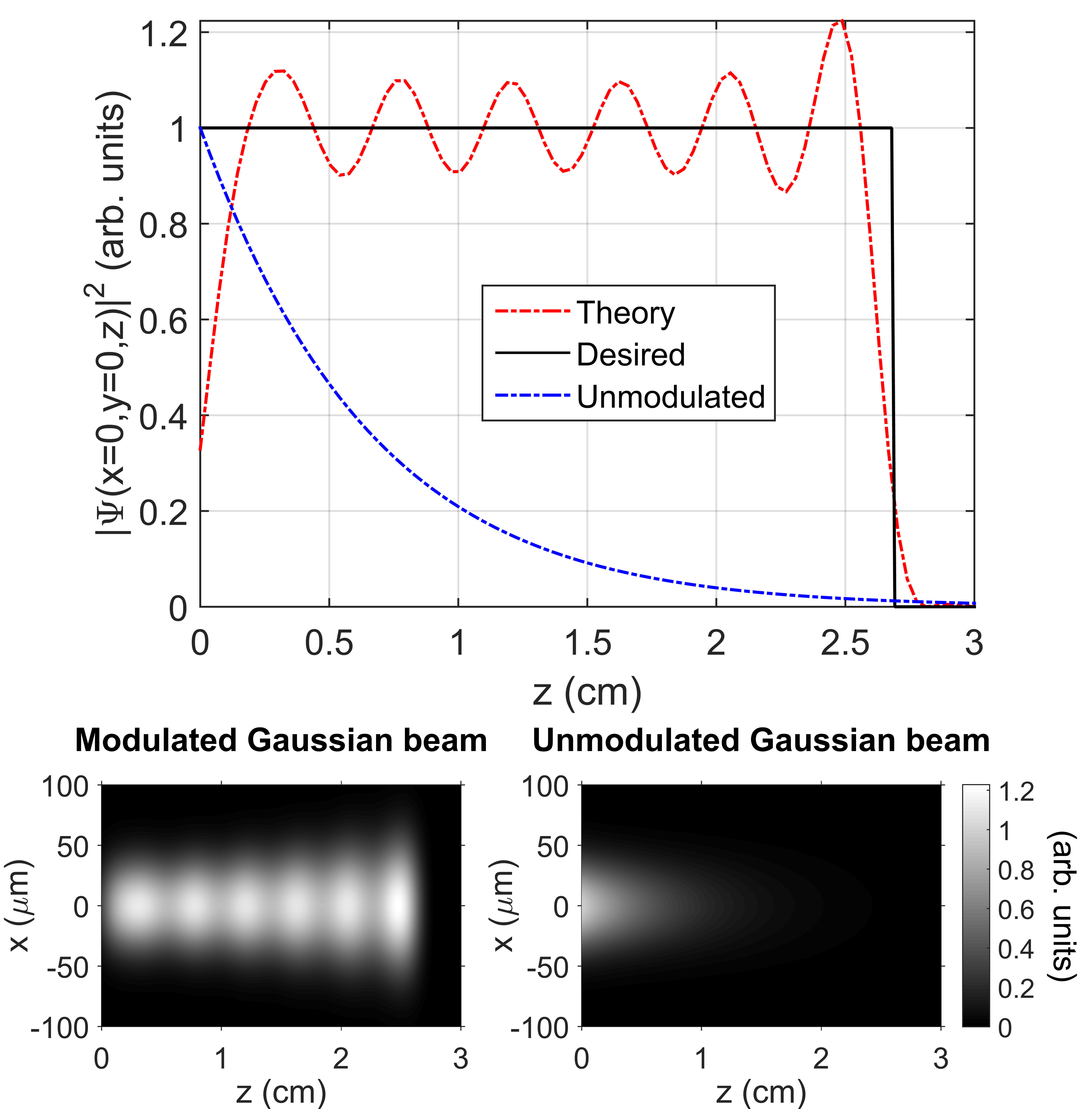}
	\caption{Constant-peak-amplitude Gaussian beam in a lossy medium. Top: desired, obtained and unmodulated on-axis longitudinal intensity patterns. Bottom: intensity patterns of the modulated and unmodulated beams in the plane $y=0$.}
	\label{perdas_1}
\end{figure}

\begin{figure} [htbp]
	\centering
	\includegraphics[width = 0.8\columnwidth]{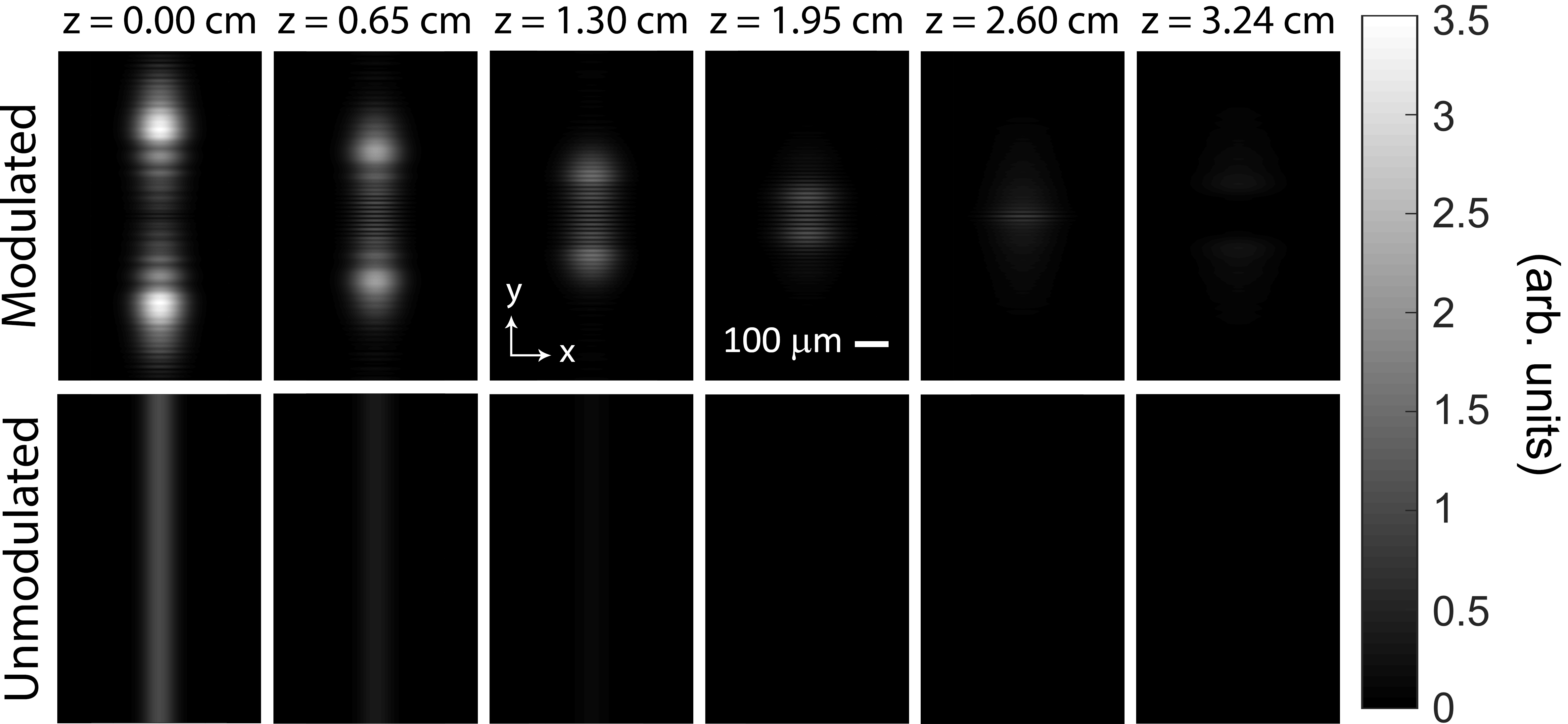}
	\caption{Constant-peak-amplitude Gaussian beam in a lossy medium: transverse intensity pattern at the some longitudinal positions for the modulated and unmodulated beams.}
	\label{perdas_2}
\end{figure}

 Fig. \ref{perdas_1} shows the desired and obtained longitudinal intensity patterns, contrasting them with the case when no modulation is applied. It also shows the intensity pattern of the beam in the plane $y=0$ with and without modulation. It is evident that the intensity of the unmodulated beam at $x=y=0$ decays rapidly due to absorption (and also due to diffraction), while the modulation makes this intensity approximately constant for a propagation distance 4 times greater than the usual penetration depth. It should be clear that this is achieved at the expense of having a sufficiently intense lateral structure, since the whole beam is still subject to the medium absorption. This is shown in Fig. \ref{perdas_2}, which depicts the transverse intensity patterns of the modulated and unmodulated beams at some propagation distances.

\par Indeed, the lateral structure has a peak intensity of approximately 3.5 times the intensity at the center of the beam in order to provide the energy necessary for the modulation. Therefore, even though the proposed method works for arbitrary losses and propagation distances, the required lateral structure may be prohibitively intense if they are excessively high, thus making the method more suitable for cases in which $L_d$ is a few times larger than the penetration depth. However, when the medium is lossless (as in sec. \ref{experiment}), this is not an issue and the intensity of the lateral structure depends only on the function $F(z)$.

\section{Experimental demonstrations}
\label{experiment}

\par
\begin{figure}[htbp]
	\centering
	\includegraphics[width=0.8\columnwidth]{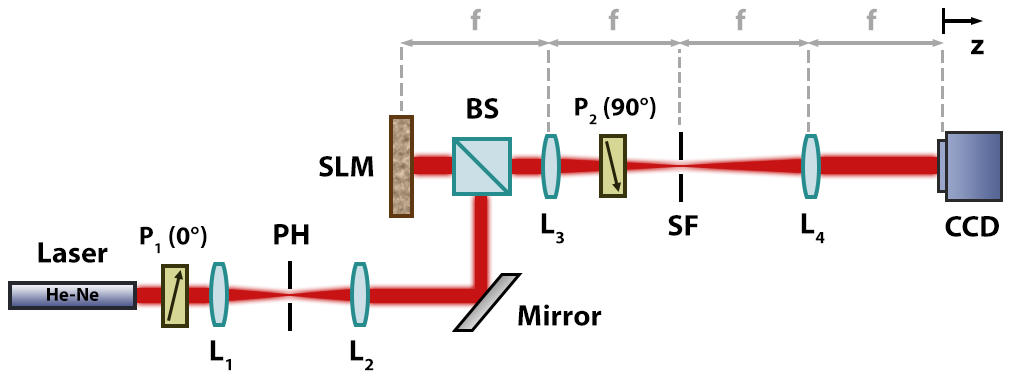}
	\caption{Experimental apparatus used for generating the desired beams. Laser: He-Ne laser; $P_1$: polarizer; PH: pinhole; $L1$ to $L4$: lenses; SLM: reflective Spatial Light Modulator, model LC-R1080 from HOLOEYE Photonics; BS: beam splitter; $P_2$: analyzer; SF: circular pupil; CCD: CCD camera model DMK 41BU02.H from The Imaging Source.}
	\label{experimental_setup}
\end{figure}

 In order to show the validity and feasibility of the proposed method, we experimentally generated some Gaussian and Airy beams modulated in the longitudinal direction by interesting functions. For simplicity, they were generated and propagated in free space. The experimental apparatus is presented in Fig. \ref{experimental_setup}.   

\par We used a reflective Spatial Light Modulator (SLM) (LC-R1080 from HOLOEYE Photonics, which has a display matrix of 1920x1200 with a pixel pitch of $8.1\,\mu\text{m}$) in the amplitude modulation mode to encode the information of the desired beam into an incident linearly-polarized Gaussian beam, which comes from a He-Ne laser ($\lambda \approx 632.8\,\text{nm}$). Before impinging on the SLM, the beam passes through a system of a pinhole (diameter of $2.5\,\mu\text{m}$), two lenses ($L_1$, with focal distance of $6.5\,\text{mm}$, and $L_2$, with focal distance of $15\,\text{cm}$) and a polarizer ($P_1$) for spatial and polarization filtering, spot expansion and collimation. The light reflected by the SLM (properly filtered by the analyzer $P_2$) is subject to a 4f spatial filtering system, in which the lens $L_3$ is placed at a focal distance $f=15\,\text{cm}$ from the SLM, and the spatial filtering mask (a band-pass circular pupil, referred to as SF in Fig. \ref{experimental_setup}) is located at its Fourier plane. The SF selects the first diffraction order, which contains the information of the desired complex field pattern. Another Fourier transform produces the field at the Fourier plane of the lens $L_4$ (focal distance of $15\,\text{cm}$) and its intensity profile is measured as a function of the longitudinal coordinate $z$ by means of a CCD camera (DMK 41BU02.H from The Imaging Source, whose display is a 1280x960 matrix of pixel size $4.65\,\mu\text{m}$).

\par The low resolutions of both the SLM and the CCD camera severely limited our choices for the parameters of $\chi(x,z)$ and $\gamma(y,z)$, as the phase and amplitude variations of the desired field at $z=0$ and the measured intensity patterns had to be well resolved respectively by the SLM and CCD camera displays. Consequently, we had to choose a paraxial $\chi(x,z)$ ($k_{x_0}=0$) and a big spot $\Delta y$, which, from Eq. \eqref{spot_paraxial}, demands a small $Q$. Thus, we chose $Q=2 \pi M/L$, the minimum value allowed by Eq. \eqref{Q_paraxial}. In order to have a reasonable approximation for $F(z)$, $M$ had to be taken as a few tens, but these choices required a big $L$ to result in spot sizes of tens of $\mu\text{m}$ that could be properly measured. However, as mentioned before, $L$ does not have to be the whole region in which the field is modelled, so we purposely increased its value without changing the intended patterns. Denoting $L_d$ the desired modulation region in each example (as in the end of sec. \ref{theory} and in sec. \ref{sec:perdas}), we adopted $L = 5L_d$ as the interval in which the Fourier coefficients $A_m$ were calculated for the modified modulation function 
\begin{equation}
F^\prime(z)=\begin{cases}
F(z)\,\text{,} &\text{for } 0\leq z\leq L_d\\
0\,\text{,} &\text{for } L_d\leq z\leq L
\end{cases}
\end{equation} 

\noindent thus allowing us to obtain the desired spot sizes without prohibitively worsening the approximation of the desired $F(z)$ for a given $M$.

\subsection{Gaussian beam with constant peak amplitude}
\label{exp_gauss}

\par The first example consists in an initially  unidimensional paraxial Gaussian beam with spot size $w_0 = 200/\sqrt{2}\,\mu\text{m}$ whose intensity was modulated to provide a constant amplitude at $x=y=0$. The parameters chosen were $L_d=0.5\,\text{m}$, $M=20$, $L=5L_d$ and $Q=2\pi M/L$, resulting in $\Delta y \approx 49.72 \mu\text{m}$. As in sec. \ref{sec:perdas}, $F(z)$ was chosen according to Eq. \eqref{F_for_gaussian}, thus compensating the natural intensity diminishing due to diffraction.

\par
\begin{figure} [htbp]
	\centering
	\includegraphics[width =0.7\columnwidth]{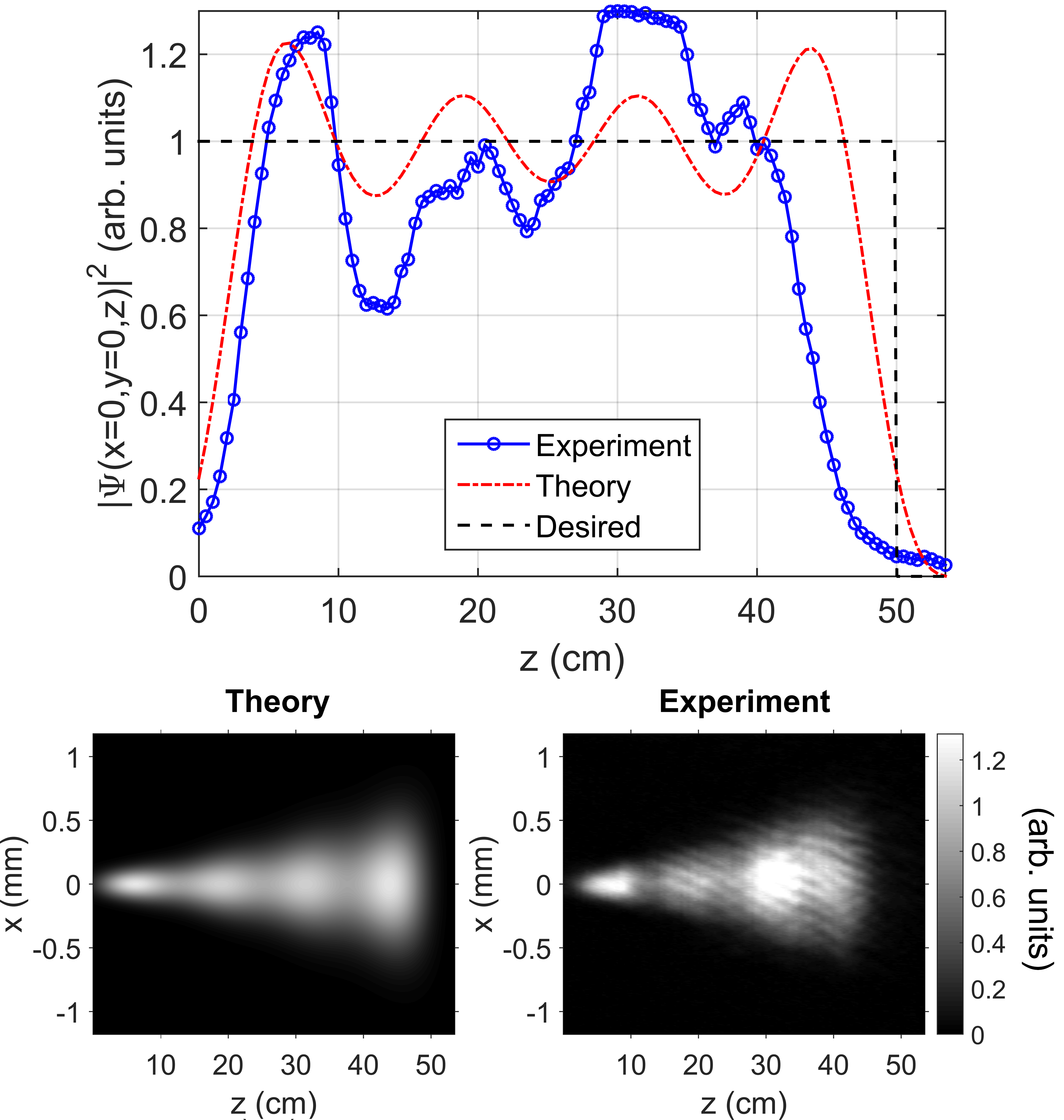}
	\caption{Constant-peak-amplitude Gaussian beam. Top: desired, theoretical and experimental on-axis longitudinal intensity patterns. Bottom: theoretical and experimental intensity patterns in the plane $y=0$.}
	\label{Gauss_exp_z}
\end{figure}

 The results obtained are shown in Figs. \ref{Gauss_exp_z} and \ref{Gauss_exp_x,z}. Fig. \ref{Gauss_exp_z} compares the experimentally obtained, theoretically expected and ideally desired on-axis intensity profiles and also shows the theoretical and experimental patterns in the plane $y=0$. Fig. \ref{Gauss_exp_x,z} presents $y=0$ cuts of the measured and expected intensity profiles, together with the captured and theoretical transverse intensity patterns at some longitudinal distances. Due to the limited number of terms retained in the truncated Fourier series for representing $F(z)$, it is no surprise that the expected on-axis intensity is not exactly constant as ideally desired, but in fact has small oscillations, as shown in Fig. \ref{Gauss_exp_z}. The measured longitudinal intensity pattern does not match exactly the expected one, but indeed follows its variations around 1 arb. units. We believe that these deviations may have been caused by an imprecise reproduction of the beam's lateral structure when it was experimentally generated. Even though we had carefully chosen a spot size that could be reasonably resolved, the pattern in the y-direction is composed of a combination of tens of cosine functions with different periods (see Eq. \eqref{psi_cos_paraxial}), which may have resulted in features that were poorly reproduced by the SLM. In addition, other imperfections in the generation may have unnoticeably introduced defects in the beam's lateral structure. Indeed, if we compare the theoretical and measured initial transverse structure of the beam in Fig. \ref{Gauss_exp_x,z} (indicated by the number 1), we can see that the generated beam extends a little less in the y-direction than what is necessary according to the theory. This lack of intensity at the beam's ends justifies why the measured on-axis intensity decreases a little earlier than desired, since it is the lateral structure that brings energy to the center, as clearly depicted by the succeeding transverse profiles shown in Fig. \ref{Gauss_exp_x,z}. Nevertheless, despite these imperfections, we see that the desired modulation was still approximately achieved and the experimental intensity patterns follow the theoretically expected profiles. 

\par For comparison, Fig. \ref{Gauss_exp_x,z} also shows the theoretical transverse intensity pattern of the same unidimensional Gaussian beam without the modulation. It is visible that its intensity at $y=0$ decreases, unlike in the modulated case.

\begin{figure} [htbp]
	\centering
	\includegraphics[width = 0.7\columnwidth]{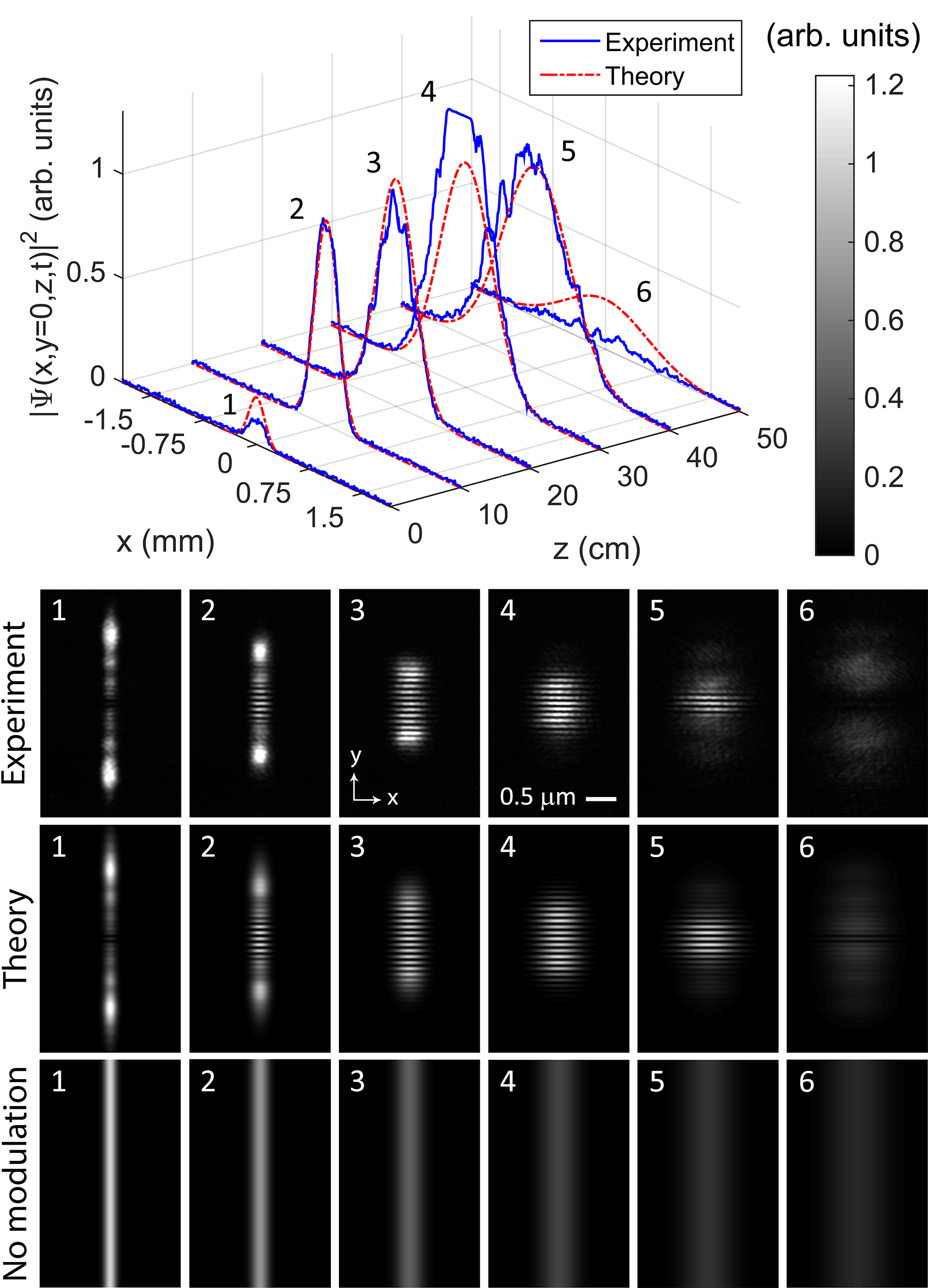}
	\caption{Constant-peak-amplitude Gaussian beam. Top: $y=0$ cuts of the measured and expected intensity patterns at some longitudinal distances. Bottom: experimental and theoretical images of the transverse intensity pattern at the same longitudinal positions. The last line also shows how the transverse intensity pattern would look like without the modulation.}
	\label{Gauss_exp_x,z}
\end{figure}

\par It is worth mentioning that the estimate for $\Delta y$ given by Eq. \eqref{spot_paraxial} is very accurate, as the experimental distance between $x=y=0$ and the first intensity null in the y-direction was measured as approximately $50\,\mu\text{m}$, as predicted. Also, the estimate of $L_y$ according to Eq. \eqref{Ly} is fairly good, resulting in $L_y\approx 1.59\,\text{mm}$, which encompasses almost the entire lateral structure. These good agreements also occur in the examples in sec. \ref{airy_exp} and \ref{airy_step}.

\par
\begin{figure} [htbp]
	\centering
	\includegraphics[width = 0.7\columnwidth]{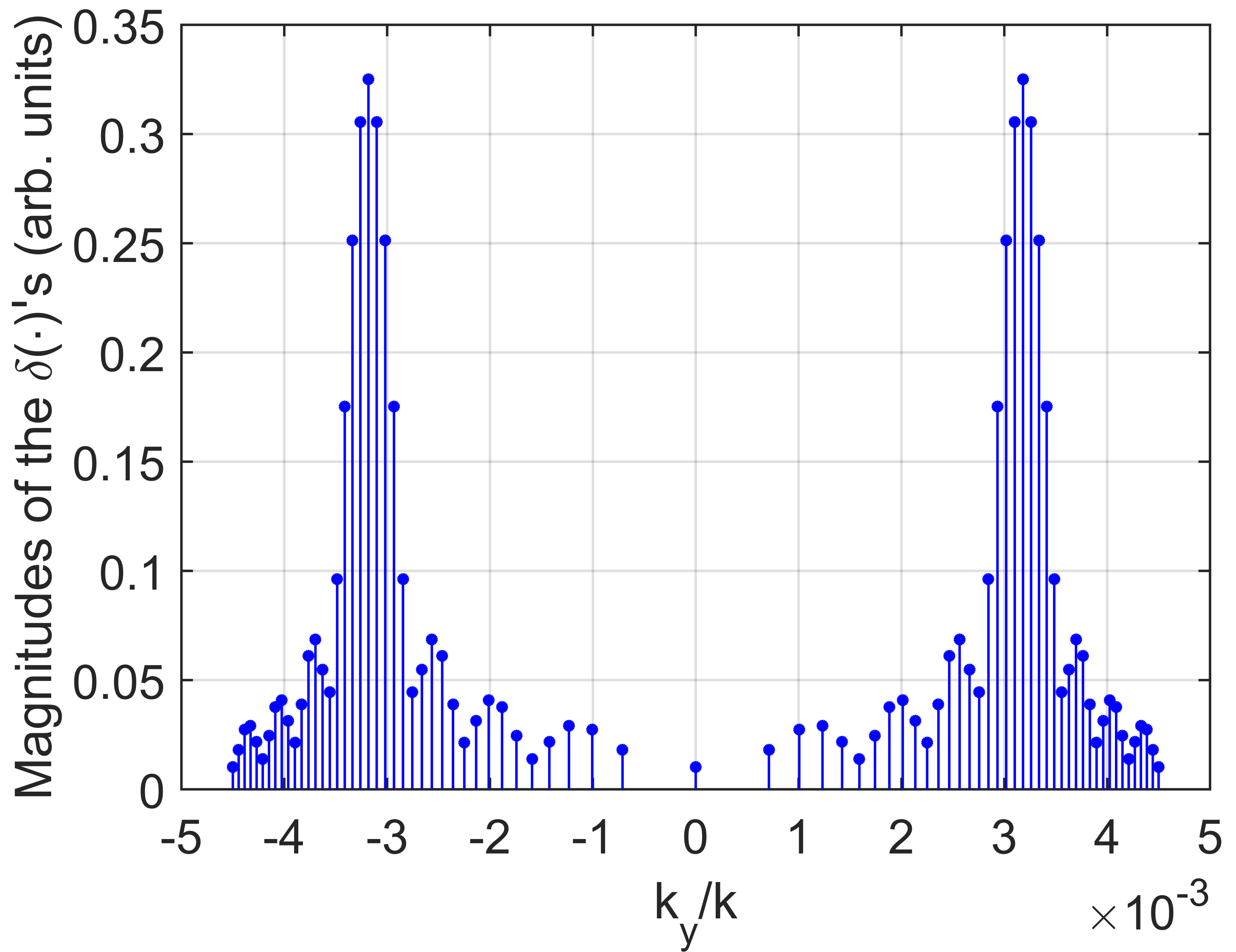}
	\caption{Constant-peak-amplitude Gaussian beam: magnitude of the $\delta(\cdot)$'s that compose $\tilde{S}_y(k_y)$.} 	
	\label{gauss_espectro}
\end{figure}

 To confirm that the proposed method indeed generated a concentrated $\tilde{S}_y(k_y)$ around $k_y=0$ and is, therefore, theoretically consistent, Fig. \ref{gauss_espectro} shows the magnitudes of the Dirac delta functions that compose $\tilde{S}_y(k_y)$. In accordance with Eq. \eqref{k_y_c}, the peaks are located at $k_y/k\approx \pm 3.18\times 10^{-3}$ and the amplitudes decrease rapidly when we move away from them.

\subsection{Airy Beam with exponentially-growing intensity}
\label{airy_exp}

\par The second example consists in an initially unidimensional Airy beam with a spot size of $x_0=100\,\mu\text{m}$ whose intensity grows exponentially. Again, the parameters chosen were $L_d = 0.5\,\text{m}$, $L = 5L_d$, $Q = 2\pi M/L$ and $M = 20$, which implied $\Delta y \approx 49.72\,\mu\text{m}$. As a reference, the expression for the envelope of an unidimensional Airy beam is \cite{airy_livro}
\begin{equation}
A_{Ai}(x,z)=\Ncal Ai\left[\frac{x}{x_0}-\frac{z^2}{4k^2 x_0^4}\right]\exp\left[i\frac{x}{x_0}\frac{z}{2kx_0^2}-\frac{i}{12}\left(\frac{z}{kx_0^2}\right)^3\right]
\end{equation}

\noindent where $\Ncal = 1/\max_x [Ai(x)]$, so that the maximum value of $|A_{Ai}(x,z)|$ is 1.

\par Since the peak values of $|A_{Ai}(x,z)|$ remain constant during propagation, the $F(z)$ chosen for an exponentially-growing intensity at $y=0$ was
\begin{equation}
F(z) = \sqrt{\exp(\frac{z}{L_d})}
\end{equation}

\par Notice that the square root is present because it is $|F(z)|^2$ that appears in the beam's intensity pattern.

\par
\begin{figure} [htbp]
	\centering
	\includegraphics[width = 0.7\columnwidth]{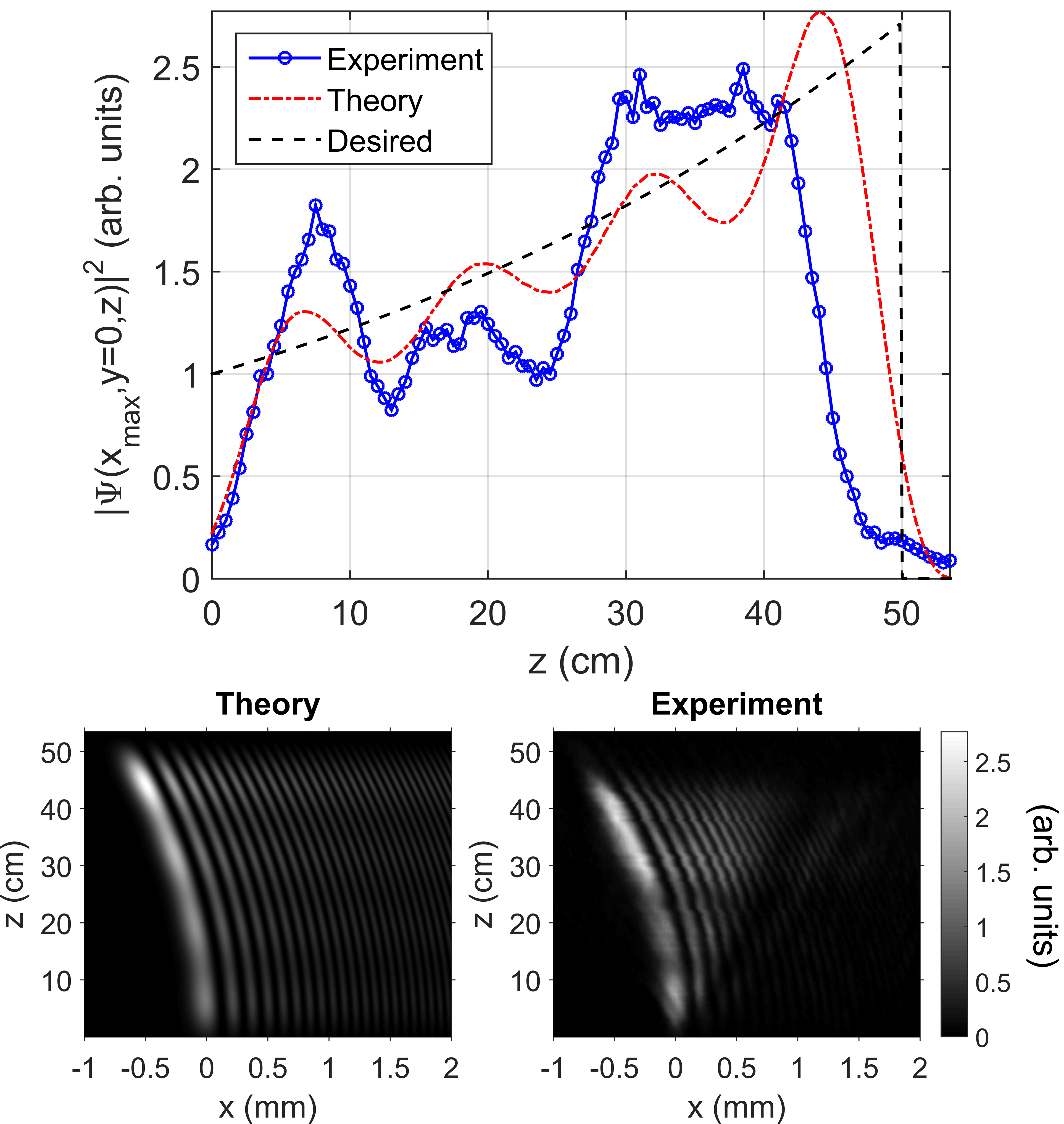}
	\caption{Airy beam with exponentially-growing intensity. Top: desired, theoretical and experimental longitudinal intensity patterns at $(x,y)=(x_{max},0)$. $x_{max}$ denotes the x-direction position of the beam's highest peak, which follows a parabolic trajectory in the $xz$-plane. Bottom: theoretical and experimental intensity patterns in the plane $y=0$.}
	\label{Airy_expon_exp_z}
\end{figure}

 Figs. \ref{Airy_expon_exp_z} and \ref{Airy_expon_exp_x,z} show the same types of results of Figs. \ref{Gauss_exp_z} and \ref{Gauss_exp_x,z} but for the current example. The characteristic parabolic trajectory of the peak of the Airy beam can be clearly seen in Fig. \ref{Airy_expon_exp_z} at the plane $y=0$. In the transverse intensity images of Fig. \ref{Airy_expon_exp_x,z}, this bending appears as increasing shifts of the peak in the positive $x$-direction.

\par Similarly to what happened in sec. \ref{exp_gauss}, the measures deviate a little from the theoretical expectations and the reasons are the same already pointed out. In addition to the experimental limitations mentioned in that section, Fig. \ref{Airy_expon_exp_x,z} clearly shows that, despite the reasonable spot size $x_0$ chosen, just a few secondary peaks of the Airy pattern in the y-direction were correctly reproduced at the initial plane. Since they are responsible for the construction of the main peak and its parabolic trajectory, we notice a little shift in its expected position as $z$ increases. Nevertheless, even though the experimental conditions for generating the modulated Airy beam were very non-optimal, the obtained results follow the theoretical predictions. 

\begin{figure} [htbp]
	\centering
	\includegraphics[width = 0.7\columnwidth]{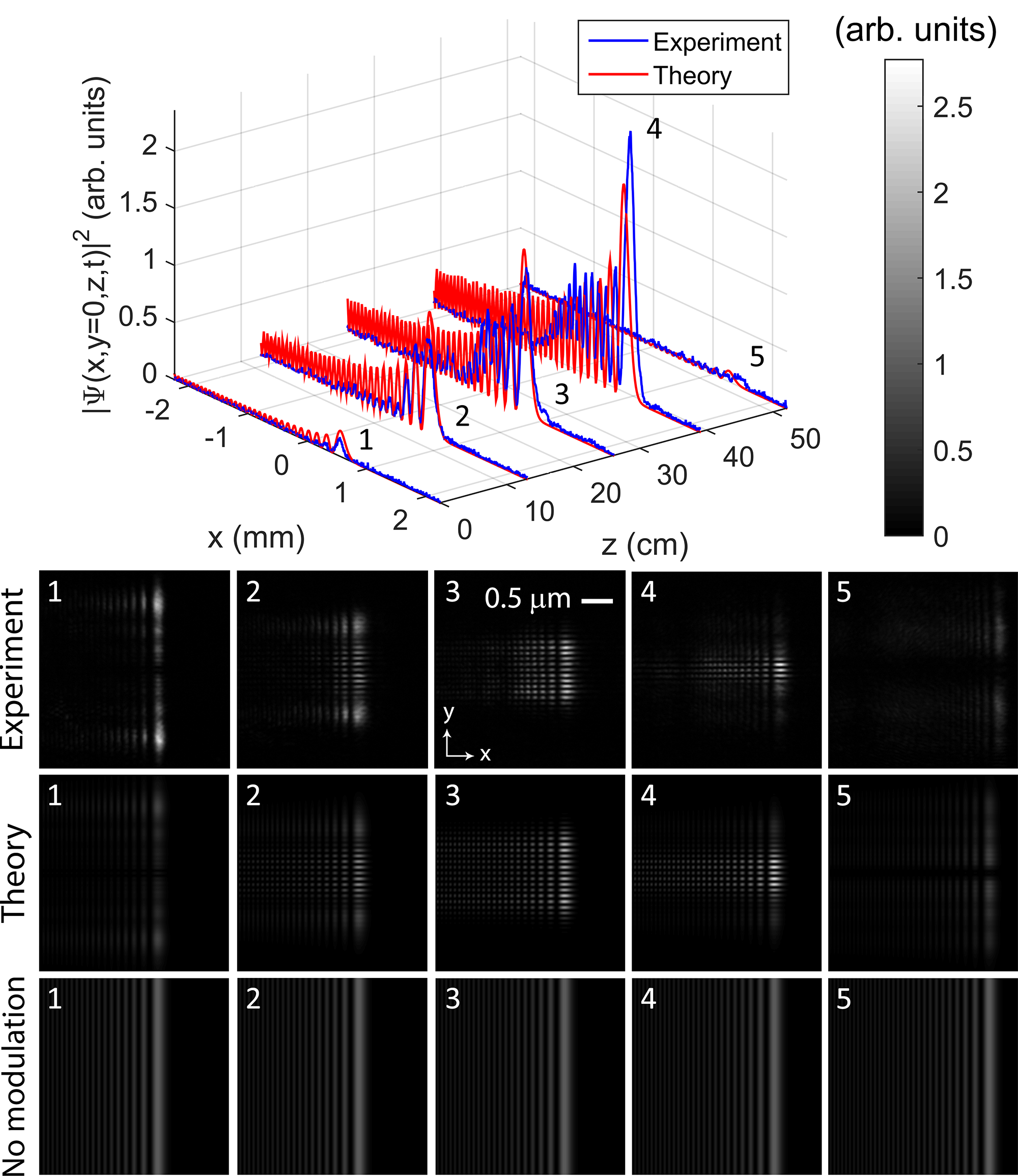}
	\caption{Airy beam with exponentially-growing intensity. Top: $y=0$ cuts of the measured and expected intensity patterns at some longitudinal distances. Bottom: experimental and theoretical images of the transverse intensity pattern at the same longitudinal positions. The last line also shows how the transverse intensity pattern would look like without the modulation.}
	\label{Airy_expon_exp_x,z}
\end{figure}

\subsection{Airy Beam with rectangular intensity pattern}
\label{airy_step}

\par The third and last example consists in an initially unidimensional Airy Beam with a rectangular intensity pattern, that is, for $y=0$ cuts, the beam is ``turned on'' within a certain interval and then disappears. In this case,
\begin{equation}
F(z)=\begin{cases}
0\,\text{,} & \text{ for } 0\leq z < L_d/3 \\
1\,\text{,} & \text{ for } L_d/3\leq z < 2L_d/3 \\
0\,\text{,} & \text{ for } 2L_d/3\leq z \leq L_d \\
\end{cases}
\end{equation} 

\par Once more, we chose $L_d = 0.5\,\text{m}$, $L = 5L_d$, $Q = 2\pi M/L$ and $x_0 = 100\, \mu\text{m}$. This time, however, $M = 40$, since the discontinuity in $F(z)$ requires a higher value of $M$ for a reasonable approximation. As a result, $\Delta y \approx 35.16\,\mu\text{m}$.

\par
\begin{figure}[htbp]
	\centering
	\includegraphics[width = 0.7\columnwidth]{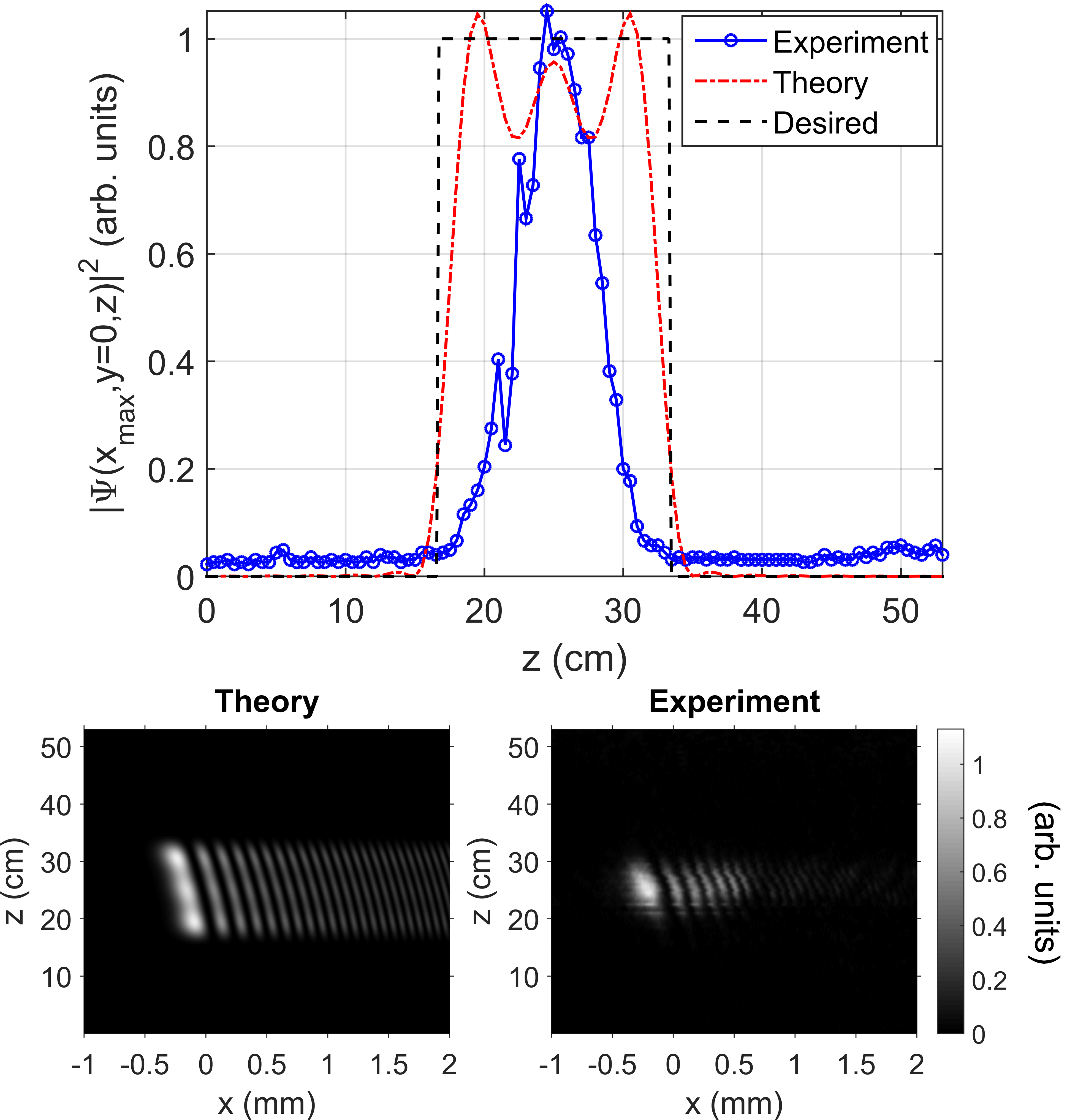}
	\caption{Airy beam with rectangular intensity pattern. Top: desired, theoretical and experimental longitudinal intensity patterns at $(x,y)=(x_{max},0)$. $x_{max}$ denotes the x-direction position of the beam's highest peak, which follows a parabolic trajectory in the $xz$-plane. Bottom: theoretical and experimental intensity patterns in the plane $y=0$.}
	\label{Airy_step_exp_z}
\end{figure}

 Figs. \ref{Airy_step_exp_z} and \ref{Airy_step_exp_x,z} show the same types of results of Figs. \ref{Gauss_exp_z} and \ref{Gauss_exp_x,z} but for the current example. Again, the results do not match exactly the theoretical predictions due to the reasons previously presented. As in the previous examples, we attribute the premature falling edge of the experimental intensity pattern to the imperfect generation of the outer end of the lateral structure of the beam. However, in this case we also see a delay in the rising edge of the profile compared to the theoretical prediction, which may also be due to an imperfect generation of the inner portion of the beam lateral structure. Nevertheless, the results illustrate that the method proposed is reasonable.

\begin{figure} [htbp]
	\centering
	\includegraphics[width = 0.7\columnwidth]{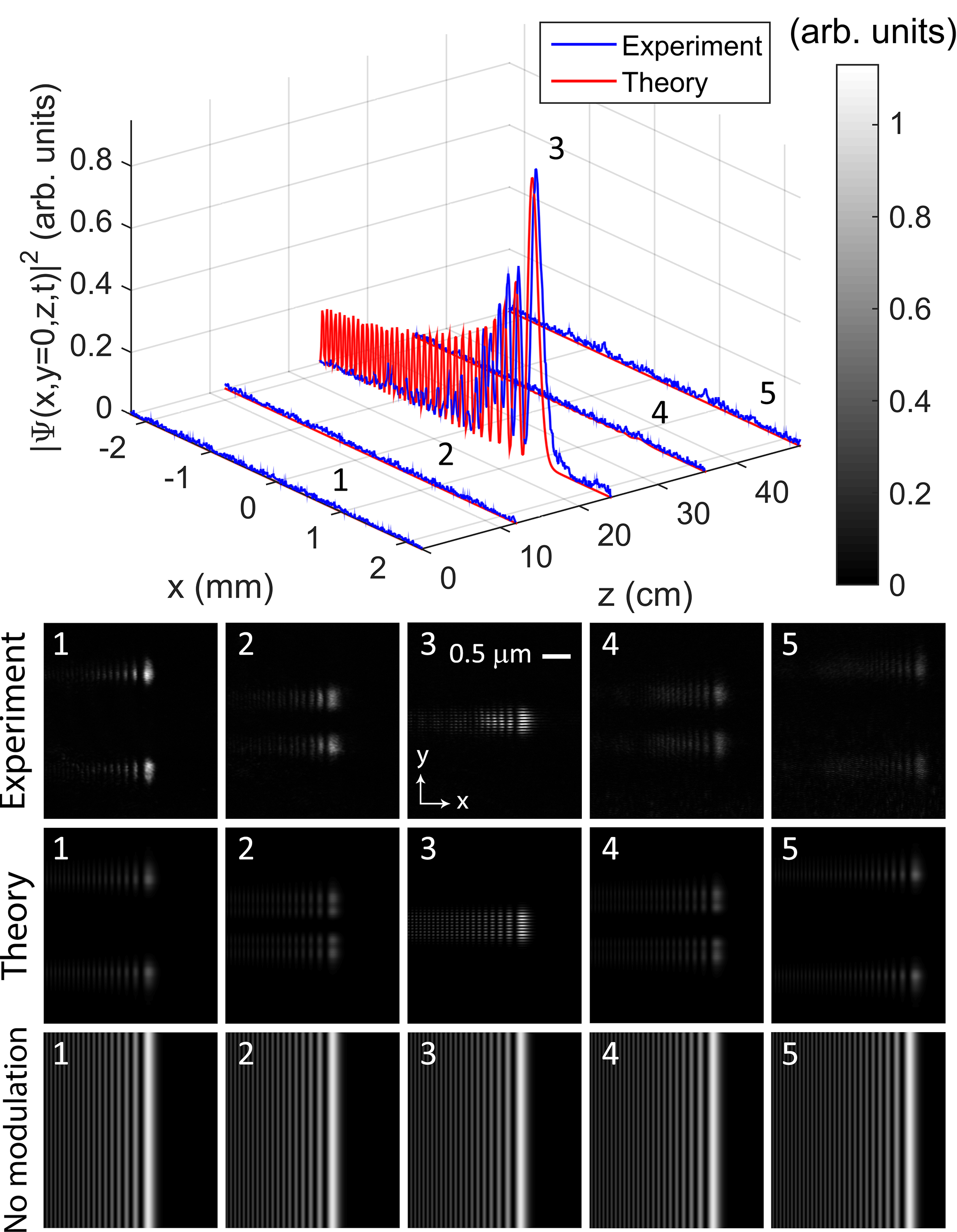}
	\caption{Airy beam with rectangular intensity pattern. Top: $y=0$ cuts of the measured and expected intensity patterns at some longitudinal distances. Bottom: experimental and theoretical images of the transverse intensity pattern at the same longitudinal positions. The last line also shows how the transverse intensity pattern would look like without the modulation.}
	\label{Airy_step_exp_x,z}
\end{figure}

\section{Conclusions}

\par In this work, we developed a method to arbitrarily modulate the longitudinal intensity pattern of (initially) unidimensional Cartesian beams with concentrated angular spectra in lossless and lossy media, which may be useful for applications such as optical tweezers and atom guiding. By writing a bidimensional beam as a product of two unidimensional beams, having the desired one depending on $x$ and $z$ and the additional one, on $y$ and $z$, and choosing appropriately the angular spectrum of the latter, we showed that it is possible to have the desired unidimensional beam structure with an arbitrary longitudinal intensity pattern in the plane $y=0$, while in the y-direction the resulting beam presents a sinusoidal-like oscillation with a spot around $y=0$ whose size can be tuned. In the case of lossy medium, the technique is able not only compensate the exponential decay caused by medium absorption, but also to result in any desired longitudinal intensity pattern.

\par Finally, a theoretical example and three sets of experimental data were presented to prove the validity and feasibility of the method. Although the measured results were not perfect due to experimental imperfections in the generation of the beams' lateral structures, the desired modulations were approximately achieved and exemplified some longitudinal intensity patterns that may be of practical interest, such as keeping the intensity of a point constant by compensating its diminishing due to diffraction, having an exponentially-growing peak intensity or selectively turning the beam ``on'' and ``off'' within desired regions.

\section{Funding}

S\~{a}o Paulo Research Foundation (FAPESP) (2015/26444-8); National Council for Scientific and Technological Development (CNPq) (304718/2016-5).

\bibliographystyle{abbrv}
\bibliography{mybib}

\end{document}